\documentstyle[prl,aps,floats,twocolumn]{revtex}
\begin{document}
\def\be{\begin{equation}}
\def\ee{\end{equation}}
\def\bear{\begin{eqnarray}}
\def\eear{\end{eqnarray}}
\def\E{{\rm e}}
\def\bearst{\begin{eqnarray*}}
\def\eearst{\end{eqnarray*}}
\def\peleven{\parbox{11cm}}
\def\peffec{\peight{\bearst\eearst}\hfill\peleven}
\def\pspace{\peight{\bearst\eearst}\hfill}
\def\ptwelve{\parbox{12cm}}
\def\peight{\parbox{8mm}}
\twocolumn[\hsize\textwidth\columnwidth\hsize\csname @twocolumnfalse\endcsname
\title
{Gravitational Clustering to All Perturbative Orders}
\author
{Elcio Abdalla$^a$ and 
Roya Mohayaee$^b$}
\address
{$^a$\it Instituto de 
F\'\i sica, USP, S\~ao Paulo\\
$^b$ HEP, ICTP, Trieste\\
}
\maketitle

\begin{abstract}

We derive the time evolution of the density contrast
to all orders of perturbation theory, 
by solving the Einstein equation for scale-invariant 
fluctuations. These fluctuations are represented by
an infinite series in inverse powers of the radial 
parameter. In addition to the standard growing modes,
we find infinitely many more new growing modes 
for open and closed universes.


\end{abstract}
\hspace{.2in}
]


The large scale structure of the universe
is believed to have grown, due to gravitational instability,
from small primordial
density fluctuations.
These fluctuations are fully characterized 
by the density contrast, 
$\delta(t,\vec{x})=\delta\rho(t,\vec{x})/\rho(t)$.
At any given time, the connected correlation
functions of the density contrast,
determine the power spectrum and the space 
distribution of the matter in the universe 
\cite{peeble,fry,goroff,frieman,bouchet}. 
The time dependence of the density 
contrast, on the other hand, 
determines if the inhomogeneities 
would indeed grow, or would just oscillate 
or decay, in the first place.

In this article,
we are primarily concerned with the time evolution of 
the density contrast. In almost all standard 
methods, this is determined by the
continuity, Euler and Poisson equations in the Newtonian regime 
\cite{peeble,weinberg,kolb,lucchin,report,report2} 
and by the Einstein equation in the relativistic era
\cite{peeble,weinberg,kolb,lucchin,report,report2}.
In the nonlinear regime, the perturbation is
carried to second order and new growing modes
containing nonlocal terms are obtained
for dust universes \cite{peeble,bouchet,martel}.
The local nonlinear terms have also been studied
to all orders for the flat universe \cite{frieman,pablo}.
A single complete treatment covering all possible cases 
and also extending to an arbitrary perturbative 
order does not exist.

Our starting point is not the 
usual equations of fluid hydrodynamics
but the Einstein equation which
naturally covers the present era too.
We expand the Einstein equation around the 
background Friedmann universe
by scale-invariant fluctuations.
These fluctuations are expressed as an infinite series 
in inverse powers of the radial parameter. 
This method, which assumes spherical symmetry, allows 
us to obtain the full expression for the growth rate
of density inhomogeneities in Newtonian, 
relativistic, zero and non-zero curvatures, 
linear and  local-nonlinear 
regimes in a single treatment. 
Thus in this way, we make major simplifications 
on current perturbative methods.
In a flat universe,
the leading growing modes are given by an infinite series 
in the matter and radiation -dominated eras.
For the open and closed universes we
find, in addition to the modes presented in the literature 
\cite{lucchin,martel}, infinitely many more new growing modes 
which are also given by infinite sums. 
In the nonlinear regime, we also find new growing 
modes for closed and open universes in the 
extreme limits of small and large times.

Contrary to the standard analysis, where the density contrast
is often taken to be a random Gaussian field, we take it to be
a scale-invariant quantity.
That the standard growing modes are contained in our results, 
is a further solid evidence that the time evolution of 
the density contrast is unaffected by its statistical properties.

Since, the inverse radial parameter now 
plays the r\^ole of 
the perturbation parameter, the order by
which the structures at different scales 
are formed is easily determined. 
Our perturbation scheme implies a 
bottom-up or a hierarchical clustering scenario
for the formation 
of structures and thus a universe dominated by cold dark matter 
\cite{lucchin}. 

To account for the inhomogeneity of space-time, we use the spherically
symmetric metric,
\bear
{\rm diag} & &\left(g_{\mu\nu}\right) =   \nonumber\\
& &\left( -1,\, {{R_p}^2(t,r)\over 1-kr^2},\, {R_p}^2(t,r)r^2,\, 
{R_p}^2(t,r)r^2 \sin^2\theta\right),\label{metric} 
\eear
where the scale factor $R_p(t,r)$ is a function of both
time and coordinate. This metric does not assume
homogeneity and is contained in the Tolman metric for
a pressure-less universe \cite{tolman}, although not restricted 
to this era.
We assume that the inhomogeneities are given by
scale-invariant fluctuations and expand the scale factor as
\be
R_p(t,r)=R(t)+\sum_{n=0}^\infty {\delta R_n(t)\over r^n}, \label{ex} 
\ee
around the background 
Friedmann-Robertson-Walker scale factor, $R(t)$.
This is a genuine perturbation of the metric
and not just a gauge mode. A solution $\zeta_\mu$ of
$\delta g_{\mu\nu}=\zeta_{\mu,\nu}+\zeta_{\nu,\mu}$ 
where $\delta g_{\mu\nu}$ is the metric perturbation, 
such that $\delta g_{0r}=0$
and $\delta g_{rr}\sim t^a r^{-n}$, for 
a general $a$ and $n$, cannot be constructed \cite{ellis}.

The time-time and radial-radial components of the
Einstein equation are expanded as in (\ref{ex})
and, on using the equation of state $P=\omega\rho$, are reduced to
the following second-order recursive differential equation \cite{elcio}:
\vfill\eject
\bear
& &\Large[\left( 2\delta\ddot R_l \delta R_s 
+(3\omega+1)\delta \dot R_l 
\delta \dot R_s
\right)r^{-(l+s)}\nonumber\\
&+&\left(\omega(3-4n-nm+2n^2)+n(m-2)+1\right)
k\nonumber\\
&-&n\left(\omega(2n-2-m)+(m-2)\right) 
r^{-2}\Large]\nonumber\\
& &
\delta R_n \delta R_m r^{-(n+m)} = 0,
\label{recurse}
\eear
where summations over the indices $n, m, l$ and $s$ are implied
and $\delta R_n$ represents $R+\delta R_0$ when $n=0$.
We have solved the above equation by a Maple program 
to very large orders for a flat universe, {\it i.e.} $k=0$ \cite{maple}.
The solutions are substituted back in the
time-time component of the Einstein equation and
a series expression for the density and subsequently the 
leading-order growing modes for the 
density contrast are found. 
The fastest growing modes can
be written in the closed forms
\be
\delta_+\sim\sum_ {n=0}^\infty 
{C}_{2n+3}{t^{(2n+2)/3}\over r^{2n+3}},\label{md}
\ee
in the matter-dominated era, $\omega=0$, and
\be
\delta_+\sim \sum_{n=0}^\infty 
{C^\prime}_{2n+3} {t^{n+1}\over r^{2n+3}}\label{rd}
\ee
in the radiation-dominated era.
We see that in addition to the usual growing 
modes, $t^{2/3}$ and $t$ in the matter and 
radiation dominated eras respectively, 
we have infinitely many 
more growing modes. The higher growing modes 
are the local nonlinear modes \cite{peeble,pablo}.
The above density contrasts also tell us that smaller 
structures enter nonlinear 
regime, $\delta>1$, faster than the larger
structures. Thus a hierarchical clustering scenario 
is expected in this scheme.

For the closed and open universes, the recursive 
equation (\ref{recurse}) is highly inhomogeneous and 
cannot be solved. However, assuming that different perturbative orders
evolve independently of each other, we find, 
in the matter-dominated era, the linearized
equation
\be
\frac{d^2{\delta R}_n}{d\psi^2} +\left( n-\frac 1{2\sinh^2\frac{\psi}2}
\right){\delta R}_n =0\quad , \label{diff-eq-open} 
\ee
where $\psi$ is real for an open and 
is imaginary, $\psi=-i\theta$, for a closed universe.
The above equation can be solved exactly. The solutions 
can be used in the Einstein equation to 
obtain the density. The 
growing modes of the density contrast are given by
\bear
\delta_+(\psi)&\sim&\frac{\sin\left(\sqrt n\psi\right)}{\sqrt n}\left[
\frac{3\sinh\psi}{\left( 1-\cosh\psi\right)^2} +\frac {2n\sinh\psi}
{1-\cosh\psi}\right] \nonumber\\
&+&\cos\left(\sqrt n\psi\right)
\left[\frac{5+\cosh\psi}{1-\cosh\psi}\right].\quad
\label{delta-plus-open} 
\eear
We see that in addition to the usual growing 
mode \cite{lucchin} which is given by $n=0$ terms 
in the above equation, 
infinitely many more growing modes exist in the linear regime.
The $n\not=0$ modes have been overlooked in the
previous analyses and can have important consequences for the
power spectrum.

In the nonlinear regime, the full inhomogeneous 
equation (\ref{recurse}) for open and 
closed universes can also 
be solved analytically at low orders and by a 
fully algebraic Maple program at higher orders \cite{elcio} 
in the limits of small and 
large times. 
At small times, the first growing modes
$\delta_+\sim t^{2/3}$ and $\delta_+\sim t$ in the 
matter and radiation-dominated eras
occur at the lowest order in the perturbation series, 
{\it i.e.} at the r-independent 
order \cite{elcio}. At higher orders the 
solution contains oscillatory 
and or decaying modes only. Therefore, 
open and closed universes behave as
a flat universe at small times in 
the linear regime, as expected.
However, this statement is not completely true. 
In a flat universe the first growing modes appear at the
third, $r^{-3}$, rather than the first, order in the perturbation 
expansion, as expressed by
equations (\ref{md}) and (\ref{rd}).
In other words, at small times 
the inhomogeneities will enter the 
nonlinear phase faster in the open and 
closed universes than in the flat universe.
Therefore, unlike common expectations, structures
can grow faster in an open 
universe than in a flat universe.

At large times, a new growing mode
arises in an open universe. 
Taking the asymptotic limit of the 
hypergeometric solutions to 
the Einstein equation \cite{elcio} we obtain
\be
\delta_+\sim {t^{3\omega/2}\over r}.
\ee
This result indicates that, at very large times,
inhomogeneities will only grow in an open universe if the
radiation pressure is non-vanishing. However, since the
baryonic matter cannot grow in the radiation-dominated
era due to its strong coupling to the radiation, the 
above mode is only relevant for the growth of
the perturbation in the non-baryonic dark matter component.

Nonlinear-nonlocal growing modes \cite{peeble} which would arise
if radial coordinates, in expression (\ref{recurse}),
were taken to be at different points,
shall be discussed in future works.

E.A thanks Conselho Federal de Desenvolvimento Cient\'\i fico e
Tecnol\'ogico (CNPq-Brazil) for financial support.


\end{document}